\documentclass[runningheads,a4paper]{llncs}

\usepackage[utf8]{inputenc}
\usepackage[OT1]{fontenc}
\usepackage[english]{babel}

\usepackage{amsmath}
\usepackage{amsfonts}
\usepackage{amssymb}
\usepackage{amssymb}
\usepackage{bm}

\usepackage[usenames,dvipsnames]{xcolor}
\usepackage{booktabs}
\usepackage{graphicx}
\usepackage{xcolor}
\usepackage{epsfig}
\usepackage{etoolbox}
\usepackage{url}
\usepackage[bookmarks]{hyperref}

\newcommand{\keywords}[1]{\par\addvspace\baselineskip
\noindent\keywordname\enspace\ignorespaces#1}

\def\pdc{prior-data conflict}

\renewcommand{\vec}[1]{{\bm #1}}

\newcommand{\uz}{^{(0)}} % upper zero
\newcommand{\un}{^{(n)}} % upper n
\newcommand{\ul}[1]{\underline{#1}}
\newcommand{\ol}[1]{\overline{#1}}

\def\yz{y\uz}
\def\yn{y\un}

\def\yzl{\ul{y}\uz}
\def\yzu{\ol{y}\uz}

\def\ynl{\ul{y}\un}
\def\ynu{\ol{y}\un}

\def\nz{n\uz}
\def\nn{n\un}

\def\nzl{\ul{n}\uz}
\def\nzu{\ol{n}\uz}

\def\PZ{I\!\!\Pi\uz}
\def\PN{I\!\!\Pi\un}
\def\MZ{\mathcal{M}\uz}
\def\MN{\mathcal{M}\un}

\def\Eta{\mathrm{H}}
\def\EZ{\mathrm{H}\uz}
\def\EN{\mathrm{H}\un}

\newcommand{\ez}{\eta_0}
\newcommand{\eo}{\eta_1}

\def\ezl{\ul{\eta}_0}
\def\ezu{\ol{\eta}_0}

\def\ezz{\eta_0\uz}
\def\ezn{\eta_0\un}
\def\eoz{\eta_1\uz}
\def\eon{\eta_1\un}

\def\eozl{\ul{\eta}_1\uz}
\def\eozu{\ol{\eta}_1\uz}

\def\czl{\ul{c}\uz}
\def\czu{\ol{c}\uz}

\def\cnu{\ol{c}\un}

\def\ezlz{{\eta_0^l}{}^{(0)}}
\def\ezuz{{\eta_0^u}{}\uz}

\def\ezln{{\eta_0^l}{}^{(n)}}
\def\ezun{{\eta_0^u}{}\un}

\begin{document}

\mainmatter  % start of an individual contribution

\title{Sets of Priors Reflecting Prior-Data Conflict and Agreement}

\titlerunning{Sets of Priors Reflecting Prior-Data Conflict and Agreement}

\author{Gero Walter\inst{1}%
%\thanks{Gero Walter was supported by the Dinalog project
%``Coordinated Advanced Maintenance and Logistics Planning for the Process Industries'' (CAMPI).}%
\and Frank P.A.\ Coolen\inst{2}}

\institute{%
School of Industrial Engineering,\\
Eindhoven University of Technology, Eindhoven, NL\\
\url{g.m.walter@tue.nl}
\and
Department of Mathematical Sciences,\\
Durham University, Durham, UK\\
\url{frank.coolen@durham.ac.uk}
}


\maketitle

\begin{abstract}
In Bayesian statistics, the choice of prior distribution is often debatable, especially 
if prior knowledge is limited or data are scarce. In imprecise probability, sets of priors
are used to accurately model and reflect prior knowledge. This has the advantage that
prior-data conflict sensitivity can be modelled:
Ranges of posterior inferences should be larger when prior and data are in conflict.
We propose a new method for generating prior sets which, in addition to prior-data 
conflict sensitivity, allows to reflect \emph{strong prior-data agreement} by decreased posterior imprecision.
\end{abstract}

\section{Introduction}

The Bayesian approach to inference \cite{2007:robert} %\cite[see, e.g.,][]{2007:robert} 
offers the advantage to combine data and prior expert knowledge in a unified reasoning process.
It combines a parametric \emph{sample model},
denoted by a conditional distribution $f(\vec{x} \mid \vartheta)$ of data $\vec{x} = (x_1, \ldots, x_n)$ given parameter $\vartheta$
with a \emph{prior distribution} $f(\vartheta)$, expressing expert opinion on $\vartheta$.
Given $\vec{x}$, the prior distribution is updated by Bayes' Rule
to obtain the \emph{posterior distribution} 
$f(\vartheta\mid\vec{x}) \propto f(\vec{x}\mid\vartheta) \cdot f(\vartheta)$.
%or the \emph{posterior predictive distribution},
%giving the distribution of further observations based on the posterior.
%
The choice of prior distribution is often debatable.
One can employ sensitivity analysis to study the effect of different prior distributions on the inferences,
as done in robust Bayesian methods
\cite{1994:berger}. %\citeNP{2005:ruggeri}, \citeNP{2000:bergerinsuaruggeri}
The method presented in this paper also uses sets of priors, with interpretation in line with
theory of imprecise probability \cite{itip,1991:walley}, considering sets of posterior distributions as the 
proper method to express the precision of probability statements themselves:
the smaller the set of posteriors, the more precise the probability statements.
%\blue{(pars linked up here)}
%A central idea in imprecise probability is thus that the magnitude of a set of distributions,
%and, in turn, the width of probability ranges based on it,
%should reflect the precision of probabilistic knowledge.
%A situation for which this relation should hold in particular is \emph{\pdc}:
This relation should hold in particular in case of \emph{\pdc}:
% from abstract esrelpaper
%A problem that can arise in Bayesian inference is called prior-data conflict:
From the viewpoint of the prior $f(\vartheta)$, the observed data $\vec{x}$ seem very surprising,
i.e., information from data is in conflict with prior assumptions \cite{2006:evans}. %\cite[see, e.g.,][]{2006:evans}.
This is most relevant when there is not enough data to largely reduce the influence of the prior on the posterior;
it is then unclear whether to put more trust to prior assumptions or to the observations,
and posterior inferences should clearly reflect this state of uncertainty.
\cite{Walter2009a} pointed out that
both precise and imprecise models based on conjugate priors can be insensitive to prior-data conflict.

For Bayesian inference based on a precise conjugate prior,
learning from data amounts to averaging between prior and data
\cite[\S~1.2.3.1]{2013:diss-gw}.
%as will be illustrated in Sect.~\ref{sec:beta-binom}.
This is the root of prior-data conflict insensitivity: 
When observed data are very different to what is assumed in the prior,
this conflict is simply averaged out and not reflected in the variance of the posterior,
giving a false sense of certainty:
A posterior with small variance indicates that we know what is going on quite precisely, but in case of \pdc\ we do not.
Prior-data conflict is reflected by increased imprecision in inferences, so more cautious probability statements,
when using carefully tailored sets of conjugate priors \cite{Walter2009a}.
One approach is to define sets of conjugate priors via sets of canonical parameters which 
ensure \pdc\ sensitivity.
\cite{Walter2009a} suggested a parameter set shape
that balances tractability and ease of elicitation
with desired inference properties.
This approach has been applied in common-cause failure modelling \cite{Troffaes2014a}
and system reliability \cite{2015:walter}.
We further refine this approach
by complementing the increased imprecision reaction to \pdc\
with further reduced imprecision if prior and data coincide especially well,
which we call \emph{strong prior-data agreement}.
These desired inference properties are achieved through a novel, more complex parameter set shape.
%In terms of the discussion at the end of \cite[\S~3.1.4]{2013:diss-gw},
%we thus sacrifice tractability to some extent %through employing a more complex parameter set shape
%in order to obtain improved model behaviour.
%\footnote{Some first results in this direction were already presented in \cite[\S~A.2]{2013:diss-gw}.}
%
%We present an idea how to define, through a novel parameter set shape,
%sets of priors such that the corresponding sets of posteriors
%show both \pdc\ sensitivity and a reaction to strong prior-data agreement.
%***motivation: clever choice of prior sets
%***emphazise that this is basically an idea for defining sets of priors***
For ease of presentation, we restrict presentation to the Beta-Binomial model, %(see Sect.~\ref{sec:beta-binom}),
the approach is generalizable to arbitrary canonical conjugate priors.
%and we will comment on this in the concluding remarks (Sect.~\ref{sec:concluding}).
Section~\ref{sec:genbayes} gives a quick overview on Bayesian inference with sets of Beta priors. 
%\cite{Walter2009a} used a parametrization 
%that allows for an intuitive interpretation of the parameters of the Beta distribution
%and clearly shows that the update step from prior to posterior
%amounts to a weighted average in the conjugate setting (Sect.~\ref{sec:genbayes}).
The new shape is defined in terms of a parametrization
recently suggested by Bickis \cite{2015:mik-isipta}
and explained in Sect.~\ref{sec:miksworld}.
%we will therefore briefly characterise this novel parametrization of canonical conjugate priors (Sect.~\ref{sec:miksworld}),
We suggest a shape in this parametrization that reacts to 
both \pdc\ and strong prior-data agreement (Sect.~\ref{sec:boatshape}).
Section~\ref{sec:concluding} discusses generalizations and potential applications.

\section{Generalized Bayesian Inference for Binary Data}
\label{sec:genbayes}

%Consider an experiment with two possible outcomes,
%\emph{success} and \emph{failure}, and success arising with probability $p$.
%The number of successes $S$ in a series of $n$ independent trials
%has a Binomial distribution with unknown parameter $p\in [0,1]$, for 
%known $n$: $S\mid p \sim \bin(n,p)$, which means
%
%Consider the Binomial sample model
%\begin{align}
%f(s\mid p) &= P(S = s \mid p) = {n \choose s} p^s (1-p)^{n-s},\quad s \in \{0, 1, \ldots, n\}\,,
%\label{eq:binompmf}
%\end{align}
%giving the probability to observe $s$ successes in a series of $n$ independent trials
%where the success probability in each trial is $p$.
%
The Binomial distribution models the probability to observe $s$ successes in $n$ independent trials
given $p$, the success probability in each trial.
In a Bayesian setting, information about $p$ is expressed by a prior distribution $f(p)$ and
updating is straightforward if one uses a \emph{conjugate} prior distribution, for which
the posterior distribution belongs to the same family as the prior, just with updated parameters.
The conjugate prior for the Binomial distribution is the Beta distribution,%
%The conjugate prior for \eqref{eq:binompmf} is the Beta distribution,%
\footnote{We denote prior parameter values by upper index~${}\uz$ and posterior parameter values, after $n$ observations,
by upper index~${}\un$.}
\begin{align}
%$f(p\mid\nz,\yz) = \frac{p^{\nz\yz-1}\, (1-p)^{\nz(1-\yz)-1}}{B(\nz\yz,\nz(1-\yz))}$.
f(p) &\propto p^{\nz\yz-1}\, (1-p)^{\nz(1-\yz)-1}\,,
\label{eq:betadensny}
\end{align}
written here in terms of the \emph{canonical} parameters %
%\footnote{\emph{Canonical} parameters are identified from rewriting the density in canonical form;
%see for example \cite[pp.~202 and 272f]{2000:bernardosmith}, or \cite[\S 1.2.3.1]{2013:diss-gw}.
%The canonical form gives a common structure to all conjugacy results in exponential families.}
$\nz > 0$ and $\yz \in (0,1)$, where $\yz$ is the prior expectation for $p$,
and $\nz$ is a pseudocount or prior strength parameter. %
%\footnote{Since $\V(p) = \frac{\yz (1-\yz)}{\nz + 1}$ is decreasing in $\nz$,
%larger $\nz$ values lead to greater concentration of probability mass around $\yz$.}
The posterior given $s$ successes in $n$ trials is a Beta distribution with updated parameters
\begin{align}
\nn &= \nz + n\,, &
\yn &= \frac{\nz}{\nz + n} \cdot \yz + \frac{n}{\nz + n} \cdot \frac{s}{n}\,.
\label{eq:nyupdate}
\end{align}
The posterior mean $\yn$ for $p$ is a weighted average of
the prior mean $\yz$ and the observed fraction of successes $s/n$,
with weights proportional to $\nz$ and $n$, respectively.
%So $\nz$ takes on the same role for the prior mean $\yz$
%as the sample size $n$ does for the observed mean $s/n$.
%Consequently, $\nz$ represents the prior strength
%and moreover can be directly interpreted as a (total) pseudocount due to the relation $\nz = \az + \bz$.
%
%The parametrization in terms of prior mean and prior strength (or pseudocount)
This averaging between prior and data is a concern if observed data differ greatly 
from what is expressed in the prior, as such conflict is averaged out
and not reflected in the posterior.

%\section{Generalized Bayesian Inference with Sets of Beta Priors}
%\label{sec:setsofbetapriors}

\cite{Walter2009a} showed that  %Walter \& Augustin (2009),
it is possible to obtain a meaningful reaction to prior-data conflict 
by using sets of priors $\MZ$ produced through parameter sets $\PZ = [\nzl, \nzu] \times [\yzl, \yzu]$.
%(a detailed discussion of different choices for $\PktZ$ is given in \citet[\S 3.1]{2013:diss-gw}.)
More generally, \cite[\S 3.1]{2013:diss-gw} describes a framework for
Bayesian inference using sets of conjugate priors based on arbitrary parameter sets $\PZ$.
%This sets of priors approach also allows to model vague and incomplete expert knowledge.
%Applying this framework to the Beta-Binomial model described above,
Here,
each prior parameter pair $(\nz, \yz) \in \PZ$ corresponds to a Beta prior,
so $\MZ$ can be taken directly as a set of Beta priors.
Alternatively, one may take the convex hull of all Beta priors with $(\nz, \yz) \in \PZ$ as $\MZ$;
$\MZ$ then consists of all finite mixtures of Beta distributions with $(\nz, \yz) \in \PZ$.
It is a modeling decision whether to take $\MZ$ as containing only Beta priors or also the mixtures.
In the first case, bounds for all inferences can be obtained by optimizing over $\PZ$.
In the second case, optimizing over $\PZ$ will only yield bounds for all inferences
that are \emph{linear functions} of $\nz$ and $\yz$,
as the linearity ensures that bounds must correspond to the extreme points of the convex set of priors,
which are the Beta priors with $(\nz, \yz) \in \PZ$.
%For inferences not linear in $\nz$ and $\yz$, only trivial bounds might exist.%
%\footnote{For example, $\V(p)$ is nonlinear in $\nz$ and $\yz$,
%and it is intuitively clear that $\V(p)$ can be larger for mixture distributions than for `pure' Beta distributions.}
In both cases, the set of posteriors $\MN$ is obtained by updating each prior in $\MZ$ according to Bayes' Rule.
This element-by-element updating can be rigorously justified
as ensuring coherence \cite[\S 2.5]{1991:walley}, and was termed ``Generalized Bayes' Rule'' by Walley \cite[\S 6.4]{1991:walley}.
In the first case, $\MN$ is a set of Beta distributions with parameters $(\nn, \yn)$,
obtained by updating $(\nz, \yz) \in \PZ$ according to \eqref{eq:nyupdate},
leading to the set of updated parameters
\begin{align}
\PN &= \Big\{ (\nn, \yn) \mid (\nz, \yz) \in \PZ = [\nzl, \nzu] \times [\yzl, \yzu] \Big\}\,.
\label{eq:paramsets}
\end{align}
In the second case, the set of Beta distributions corresponding to $(\nn, \yn) \in \PN$
forms the extreme points of the convex set of posteriors $\MN$,
such that, just like $\MZ$, $\MN$ can be described as a set of all finite mixtures of Beta distributions
with $(\nn, \yn) \in \PN$, see \cite[pp.~56f]{2013:diss-gw}.

$\MN$ forms the basis for all inferences,
leading to probability \emph{ranges} obtained by minimizing and maximizing over $\MN$.
For example, %in the Beta-Binomial model %as discussed in Section~\ref{sec:beta-binom},
the posterior predictive probability %$[\lpr,\upr]$
for the event that a future single draw is a success is equal to $\yn$;
for an imprecise model $\MZ$ based on $\PZ$,
the lower and upper probability are
%lower and upper probabilities are
\begin{align*}
\inf_{\PN} \yn &= \inf_{\PZ} \frac{\nz\yz + s}{\nz +n} & &\text{and} &
\sup_{\PN} \yn &= \sup_{\PZ} \frac{\nz\yz + s}{\nz +n}\,.
\end{align*}
The relation between $\PZ$ and $\MZ$, as well as between $\PN$ and $\MN$,
allows to characterize model properties through properties of $\PZ$ and $\PN$,
as is done in \cite[\S 3.1.2 -- 3.1.4]{2013:diss-gw}.
%Specific imprecise probability models are obtained by certain choices of $\PZ$.
The well-known Imprecise Dirichlet Model \cite{1996:walley::idm}
corresponds to a choice of $\PZ = \nz \times (\yzl, \yzu)$ where $(\yzl, \yzu) = (0,1)$.
The model proposed by \cite{2005:quaeghebeurcooman} generally assumes $\PZ = \nz \times [\yzl, \yzu]$,
and was shown to be insensitive to prior-data conflict by \cite{Walter2009a},
who proposed parameter sets $\PZ = [\nzl, \nzu] \times [\yzl, \yzu]$ instead.
Indeed, for $\PZ = \nz \times [\yzl, \yzu]$, we get $\PN = \nn \times [\ynl, \ynu]$, where
%\begin{align*}
%\ynl &= \frac{\nz\yzl + s}{\nz + n} & &\text{and} &
%\ynu &= \frac{\nz\yzu + s}{\nz + n}\,.
%\end{align*}
%$\ynl = \frac{\nz\yzl + s}{\nz + n}$ and $\ynu = \frac{\nz\yzu + s}{\nz + n}$.
$\ynl = (\nz\yzl + s)/(\nz + n)$ and $\ynu = (\nz\yzu + s)/(\nz + n)$.
The posterior imprecision in the $y$ dimension, denoted by $\Delta_y(\PN)$, is then
\begin{align*}
%\label{eq:deltay}
\Delta_y(\PN) &= \ynu - \ynl = \frac{\nz (\yzu - \yzl)}{\nz +n}\,,
\end{align*}
and so the same for any fixed $n$, independent of $s$.
In contrast, parameter sets $\PZ = [\nzl, \nzu] \times [\yzl, \yzu]$
provide prior-data conflict sensitivity, since
%\footnote{See \cite{Walter2009a} or \cite[\S 3.1.4]{2013:diss-gw} for more details.}
\begin{align*}
%\label{eq:deltay2}
\Delta_y(\PN) &= \frac{\nzu (\yzu - \yzl)}{\nzu + n} %\nonumber\\
               + \inf_{\yz \in [\yzl,\yzu]} |s/n - \yz| \frac{n (\nzu - \nzl)}{(\nzl + n)(\nzu + n)}\,.
\end{align*}
The shape of $\PZ$ poses a trade-off \cite[\S 3.1.4]{2013:diss-gw}:
Less complex shapes are easy to handle and lead to tractable models,
but will offer less flexibility in expressing prior information
and may have undesired inference properties.
In contrast, more complex shapes may allow for more sophisticated model behaviour
at the cost of more involved handling.

\section{A Novel Parametrization for Beta Priors}
\label{sec:miksworld}

\begin{figure}  %trim=l b r t
\centering
%\fbox{%
%\includegraphics[trim = 16mm 48mm 25mm 62mm, clip, width=\textwidth]{R/boatshape-vertical2}%
\includegraphics[trim = 2mm 6mm 10mm 20mm, clip, width=0.8\textwidth]{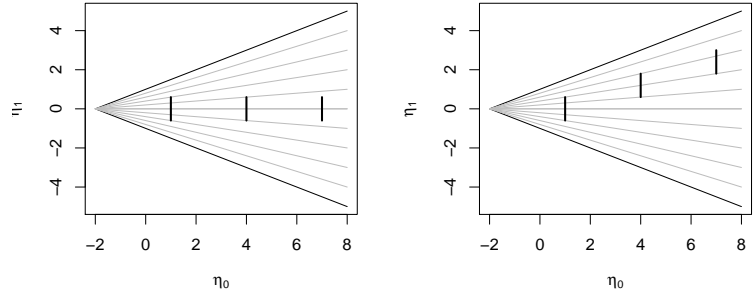}%
%}
\caption[Line segment parameter set $\EZ$ %
and respective posterior sets for $s/n=0.5$ and $s/n=0.9$.]%
%{Bounds for the domain of $\eta_0$ and $\eta_1$ for the Beta-Binomial model (black),
%with rays of constant expectation for $y_c = \{0.1,0.2,\ldots,0.9\}$ (grey).}
{Prior parameter set $\EZ = \ezz \times [\text{\underline{$\eta$}${}_1\uz$},\eozu]$
and respective posterior sets $\EN$ for $s/n=0.5$ (left) and $s/n=0.9$ (right).
Bounds for the domain $\Eta$ are in black, with rays of constant expectation for $y_c = \{0.1,0.2,\ldots,0.9\}$ in grey. 
Note that all sets have the same size, imprecision decreasing only through their position on the $\eta_0$ axis.}
\label{fig:boatshape-vertical}
\end{figure}

A conjugate Beta prior is updated by a shift in the parameter space,
given by rewriting \eqref{eq:nyupdate}:
\begin{align*}
\nz &\mapsto \nz + n\,, &
\yz &\mapsto %\frac{\nz}{\nz + n} \cdot \yz + \frac{n}{\nz + n} \cdot \frac{s}{n} = 
             \yz + \frac{s - n \yz}{\nz+n}\,.
\end{align*}
The shift for the $n$ coordinate is the same for all elements $(\nz,\yz)$ of $\PZ$. 
The shift in the $y$ coordinate depends on $\nz$, $n$, $s$, and the location of $\yz$ itself
(in fact, how far $\yz$ is from $s/n$).
The shape of $\PZ$ changes during the update step to $\PN$, %for understanding of the update step in generalized Bayesian inference,
the effects on posterior inferences may be difficult to grasp.
To isolate the influence of a set shape,
%we now consider a different parametrization of the prior
%where each coordinate has the same shift in updating.
%The updating of a parameter set
%then corresponds to a shift of the entire set within the parameter space.
%This parametrization has recently been developed by Bickis \cite{2015:mik-isipta}. %Mi\c{k}elis Bickis
we consider a recently proposed parametrization \cite{2015:mik-isipta},
where each coordinate has the same shift in updating,
such that updating a prior set corresponds to a shift of the entire set.
In this novel parametrization, a conjugate prior is represented by a coordinate $(\ezz,\eoz)$,
related to $(\nz, \yz)$ by\vspace*{-0.5ex}
%\begin{equation}
\begin{align}
\label{eq:trafotony}
%\begin{aligned}
\nz &= \ezz + 2\,, &
\yz &= \frac{\eoz}{\ezz + 2} + \frac{1}{2}\,.
%\end{aligned}
%\end{equation}
\end{align}
The domain of $\eta_0$ and $\eta_1$ in case of the Beta-Binomial model is
\begin{align}
\label{eq:eta-domain}
\Eta &= \Big\{ (\eta_0,\eta_1) \Big| \eta_0 > -2,\ |\eta_1| < \frac{1}{2}(\eta_0 + 2) \Big.\Big\}\,,
\end{align}
the Bayes update step in terms of $\eta_0$ and $\eta_1$ is given by
\begin{equation}
\label{eq:eta-update}
\begin{aligned}
\eta_0\un &= \eta_0\uz + n\,, & 
\eta_1\un &= \eta_1\uz + \frac{1}{2}(s - (n-s)) = \eta_1\uz + s - \frac{n}{2}\,.
\end{aligned}
\end{equation}
%where $s$ is again the number of successes in the $n$ Bernoulli trials.
Each success thus
leads to a step of $1$ in the $\eta_0$ direction and of $+\frac{1}{2}$ in the $\eta_1$ direction,
while each failure
leads to a step of $1$ in the $\eta_0$ direction and of $-\frac{1}{2}$ in the $\eta_1$ direction. 
While $\yz$ had the convenient property of being equal to
the prior expectation for $p$, %of the mean sample statistic $s/n$,
$\eta_1$ is only slightly more difficult to interpret.
From \eqref{eq:trafotony} we can derive that points $(\eta_0,\eta_1) \in \Eta$
on rays emanating from the coordinate $(-2,0)$,
i.e., coordinates satifying
%\begin{align}
%\label{eq:raysofconstantexpectation}
$\eta_1 = (\eta_0 + 2)(y_c - 1/2)$,
%\end{align}
will have a constant expectation of $y_c$.
The domain $\Eta$, and these \emph{rays of constant expectation} emanating from the coordinate $(-2,0)$,
can be seen in Fig.~\ref{fig:boatshape-vertical}.

In the parametrization in terms of $(\nz,\yz)$,
posterior inferences based on $\yn$ become less imprecise with increasing $n$
because $\Delta_y(\PN) \to 0$ for $n \to \infty$.
%In the domain $\Eta$, %as depicted in Figure~\ref{fig:boatshape-domain},
%instead the rays of constant expectation fan out for growing $n$,
%while a parameter set will retain its size after updating.
In the domain $\Eta$, parameter sets do not change size during update,
but the rays of constant expectation fan out for increasing $n$.
%Increased precision in a posterior parameter set $\EN$,
%which is just its prior counterpart $\EZ$ shifted to the right,
The more $\EN$ is located to the right,
the fewer rays of constant expectation it intercepts,
and so imprecision decreases.
Imprecision in terms of $\yn$
can thus be imagined as the size of the `shadow' that a set $\EN$ casts
given a light source in $(-2,0)$. %(the point from which the rays of constant expectation emanate).
The smaller this shadow, the less imprecise the inferences.
%In terms of the $(\nz, \yz)$ parametrization, we will denote the bounds of this shadow by $\ynl$ and $\ynu$:
Denoting the bounds of this shadow by\vspace*{-0.5ex}
\begin{align*}
\ynl_\Eta &:= \min_{(\ezn,\eon) \in \EN} \frac{\eon}{\ezn+2} + \frac{1}{2}\,, &
\ynu_\Eta &:= \max_{(\ezn,\eon) \in \EN} \frac{\eon}{\ezn+2} + \frac{1}{2}\,,
\end{align*}
we call the $\ez$ coordinate of $\arg\min_{(\eta_0,\eta_1) \in \EN} \yn$ and $\arg\max_{(\eta_0,\eta_1) \in \EN} \yn$
the \emph{lower} and \emph{upper touchpoint} of $\EN$ responsible for the shadow $[\ynl_\Eta, \ynu_\Eta]$.
Mutatis mutandis, the same definitions can be made for the prior set $\EZ$.
Due to the fanning out of rays, most shapes for $\EZ$ will lead to decreasing imprecision for increasing $n$.
For example, models with $\PZ = \nz \times [\yzl, \yzu]$
are represented by a line segment $\EZ = \ezz \times [\eozl,\eozu]$,
%such that the lower and upper posterior touchpoint is at $\ezn$, for any $s$ and $n$.
%Defining $\eonl$ and $\eonu$ as the updated versions of $\eozl$ and $\eozu$, respectively,
%we get $\eonu-\eonl = \eozu-\eozl$,
and imprecision decreases because a line segment of fixed size
will cast a smaller shadow when further to the right,
as illustrated in Fig.~\ref{fig:boatshape-vertical}.

For \pdc\ sensitivity, we need sets $\EZ$ that cover a range of $\eta_0$ values,
just like sets $\PZ$ with a range of $\nz$ values are necessary to ensure this property.
A set $\EZ$ that is elongated along a certain ray of constant expectation
will behave similar to a rectangular $\PZ$. %Sect.~\ref{sec:genbayes}.
When shifted along its ray of constant expectation,
imprecision will be reduced as the shadow of $\EZ$ will become smaller just as described above for line segments.
When $\EZ$ is instead shifted away from its ray of constant expectation,
imprecision will increase, as a prolonged shape that is now turned away from its ray 
will cast a larger shadow.
% (see Fig.~\ref{fig:boatshape-posterior-mik} below).%
%\footnote{This will become clear from the depiction of boatshape sets in Figure~\ref{fig:boatshape-posterior-mik}.} 
%A set $\EZ$ allowing for more precison in case of strong prior-data agreement
%must thus be able to cast a smaller shadow if the update shift is along the direction of its ray.
%of $s/n$ according the information,

\section{The Boatshape}
\label{sec:boatshape}

The shape for $\EZ$ that we suggest to obtain both \pdc\ sensitivity and reduced imprecision
in case of strong prior-data agreement looks like a boat with a transom stern
(see Fig.~\ref{fig:boatshape-posterior-mik} below).
The curvature along its length in the direction of its constant rays of expectation
leads to smaller $\Delta_y(\PN)$ as compared to a rectangular $\PZ$ with the same prior range $\Delta_y(\PZ)$,
see Fig.~\ref{fig:boatshape-posterior-normal}.
The strong prior-data agreement effect is realized
through the touchpoints determining $\ynl_\Eta$ and $\ynu_\Eta$
moving along the shape during updating,
%being attained at higher values of $\eta_0$ for larger $n$,
see Sect.~\ref{sec:spda-property}.
This is advantageous since the spread of the Beta posteriors is determined by $\eta_0 = \nz - 2$.
In case of strong prior-data agreement, variances in the `critical' distributions
at the boundary of the posterior expectation interval $[\ynl_\Eta,\ynu_\Eta]$ will thus be lower
leading to reduced imprecision.
%In particular, credible intervals for these `critical' distributions will be shorter, 
%such that the union of all credible intervals for Beta distributions with $(\ezn, \eon) \in \EN$,
%taken as a posterior inference reflecting uncertainty due to both vague prior knowledge and stochasticity, will be %shorter.
%
%In the remainder of this section, 
%we will suggest a parametrization for a curved shape in the $(\eta_0, \eta_1)$ space.
%The definition, along with some graphical examples, is given in Sect.~\ref{sec:basicdefboat},
%and we discuss a number of technical results for this shape in subsequent sections.

\vspace{-0.3cm}

\subsection{Basic Definition}
\label{sec:basicdefboat}
\begin{figure}  %trim=l b r t
\centering
%\fbox{%
%\includegraphics[trim = 20mm 35mm 30mm 45mm, clip, width=\textwidth]{R/boatshape-posterior-mik}%
\includegraphics[width=0.8\textwidth]{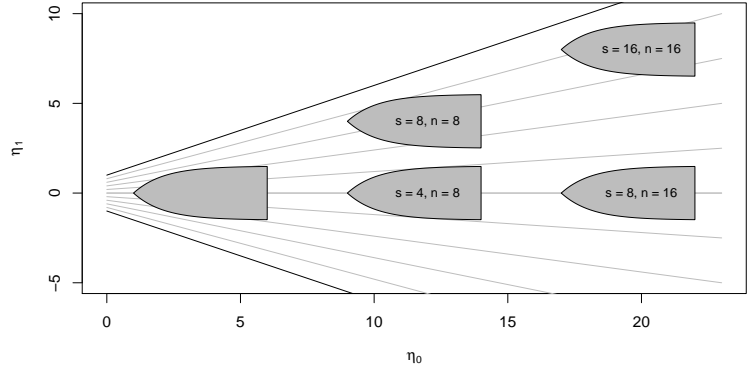}%
%}
\caption[Boatshape prior and posterior sets for data in accordance and in conflict with the prior set.]%
{Boatshape prior and posterior sets for data in accordance and in conflict with the prior set.
The parameters for the prior set are $\ezl=1$, $\ezu=6$, $a=1.5$, and $b=0.9$.
While the posterior sets for $\frac{s}{n}=0.5$ move along the ray for $y_c=0.5$,
the posterior sets for $\frac{s}{n}=1$ are shifted away from the ray for $y_c=0.5$,
resulting in increased posterior imprecision.
Note that lower and upper touchpoints are in the middle of the contour
for the prior set and the posterior sets resulting for data $\frac{s}{n}=0.5$,
while the lower touchpoint is at the end for the posterior sets for data $\frac{s}{n}=1$.}
%(See also Figure~\ref{fig:boatshape-posterior-normal}).}
\label{fig:boatshape-posterior-mik}
\end{figure}

\begin{figure}  %trim=l b r t
\centering
%\fbox{%
%\includegraphics[trim = 15mm 45mm 25mm 60mm, clip, width=\textwidth]{R/boatshape-posterior-normal}%
\includegraphics[width=0.8\textwidth]{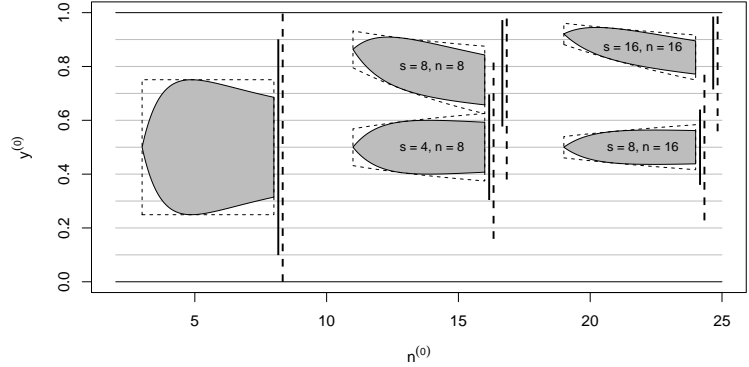}%
%}
\caption[Boatshape prior and posterior sets from Fig.~\ref{fig:boatshape-posterior-mik} in the $(\nz,\yz)$ parametrization.]%
{Boatshape prior and posterior sets from Fig.~\ref{fig:boatshape-posterior-mik} in the $(\nz,\yz)$ parametrization.
The rectangular prior set with the same range for $\yz$ as the prior boatshape set
and the corresponding posterior sets are drawn with dashed lines.
Unions of symmetric credibility intervals ($\gamma=0.5$) are drawn as vertical bars. 
Note that all posterior boatshape sets have shorter $\yn$ ranges than their corresponding posterior rectangle sets,
and boatshape credibility regions are especially short for posterior sets where $\frac{s}{n}=0.5$.}
\label{fig:boatshape-posterior-normal}
\end{figure}

\vspace{-0.3cm}
%We now present a parametrization for such a boat-shaped parameter set $\EZ$.
We suggest an exponential function for the contours of a boat-shaped parameter set $\EZ$. %
%\footnote{\blue{Other functional forms would lead to similar results.***}}
We first restrict discussion on prior sets
that are symmetric to the $\eta_0$ axis, i.e., centered around $y_c = 0.5$. 
%a fraction of successes of $\frac{s}{n} = \frac{1}{2}$ as the most probable.
Sets $\EZ$ with central ray $y_c \neq 0.5$ can be obtained
by rotating the set around $(\eta_0, \eta_1) = (-2,0)$ such that $y_c$ forms the axis of symmetry.
Results for sets with $y_c = 0.5$ generalize straightforwardly to the case $y_c \neq 0.5$;
an example is given in Fig.~\ref{fig:ppp2}.
%
%For the contours of $\EZ$, we suggest an exponential function as the functional form,
%where the `prow' of the set is located at $(\ezl, 0)$.
The lower and the upper contour functions are defined as
%\begin{align*}
%\czl(\eta_0) &= -a \left( 1 - e^{-b(\eta_0 - \ezl)} \right)\,, \\
%\czu(\eta_0) &= \phantom{-}%
%                 a \left( 1 - e^{-b(\eta_0 - \ezl)} \right)\,, 
%\end{align*}
\begin{align}
\czl(\eta_0) &= -a \left( 1 - e^{-b(\eta_0 - \ezl)} \right)\,, &
\czu(\eta_0) &=  a \left( 1 - e^{-b(\eta_0 - \ezl)} \right)\,, 
\label{eq:basiccontours}
\end{align}
where $a > 0$ and $b > 0$ are parameters controlling the shape of $\EZ$, which is defined as
%We will also need the respective derivations with respect to $\eta_0$, given by
%\begin{align*}
%\frac{d}{d\eta_0} \czl(\eta_0) &= -ab e^{-b(\eta_0 - \ezl)}\,, \\
%\frac{d}{d\eta_0} \czu(\eta_0) &= \phantom{-}%
%                                   ab e^{-b(\eta_0 - \ezl)}\,.
%\end{align*}
%Given parameters $\ezl$, $\ezu$, $a$, and $b$, $\EZ$ is defined as
\begin{align}
\label{eq:basicset}
\EZ =
\{(\eta_0,\eta_1) \colon \ezl \le \eta_0 \le \ezu, \czl(\eta_0) \le  \eta_1 \le \czu(\eta_0) \}\,.
\end{align}
A prior boatshape set,
together with corresponding posterior sets for different observations,
is shown in Fig.~\ref{fig:boatshape-posterior-mik}.
The same prior and posterior sets in terms of $(\nz, \yz)$ are depicted in Fig.~\ref{fig:boatshape-posterior-normal}.

%We have as yet no appealing formal description for the role of the parameters $a$ and $b$. %that,
%togeher with $\ezl$ and $\ezu$, define the boat-set (see Section~\ref{sec:basicdefboat} below).
The parameter $a$ determines the half-width of the set;
the size in the $\eta_1$ dimension would be $2a$ if $\ezu \to \infty$.
Parameter $b$ determines the `bulkyness' of the shape.
Together with $\ezl$, $a$ and $b$ determine %the prior interval for the expected success probability
$[\yzl_\Eta, \yzu_\Eta]$.
Decreasing $\ezl$, or increasing $a$ or $b$, leads to a wider $[\yzl_\Eta, \yzu_\Eta]$.
%For fixed $\ezl$ and $a$, increasing $b$ leads to a wider prior expectation interval.
%For $[\yzl, \yzu]$, the choice of $\ezu$ is irrelevant.%
$\ezu$ plays only a role in determining when the `unhappy learning' phase starts
(see end of Sect.~\ref{sec:generalupdate}).

We see from the prior set in Fig.~\ref{fig:boatshape-posterior-normal}
that the lower and the upper bound for $\yz$
is attained in the middle of the set contour.
To determine $\yzl_\Eta$ and $\yzu_\Eta$, we need to find the corresponding
touchpoints $\ezlz$ and $\ezuz$ %$\eta_0^{l(0)}$ and $\eta_0^{u(0)}$
by identifying the rays of constant expectation %\eqref{eq:raysofconstantexpectation}
that are tangents to $\EZ$ and then solving for $\eta_0$.
Since $\EZ$ is symmetric to the $\eta_0$ axis, we have $\ezlz = \ezuz$ 
and we will determine $\ezuz$ by considering the upper contour tangent.
We get
\begin{align}
%a - a \big(1 + b(\eta_0^u + 2)\big) e^{-b(\eta_0^u - \ezl)} &\stackrel{!}{=} 0
%& &\Longleftrightarrow &
1 + b(\ezuz + 2) &\stackrel{!}{=} e^{b(\ezuz - \ezl)}\,.
\label{eq:eta0uprior}
\end{align}
This equation only has one solution for $\ezuz > \ezl$
that is, however, not available in closed form.
Generally, the nearer $\ezuz$ is to $\ezl$, the larger $\frac{d}{d\eta_0} \czu(\ezuz)$,
such that $\yzu_\Eta$ %the upper expected value for the prior set
is further away from $\frac{1}{2}$.

\vspace{-0.3cm}

\subsection{Strong Prior-Data Agreement Property}
\label{sec:spda-property}
\begin{figure}  %trim=l b r t
\centering
%\fbox{
%\includegraphics[trim = 48mm 35mm 49mm 47mm, clip, width=0.8\textwidth]{R/prior-vs-posterior-eta0u.eps}%
\includegraphics[trim = 1mm 5mm 10mm 20mm, clip, width=0.8\textwidth]{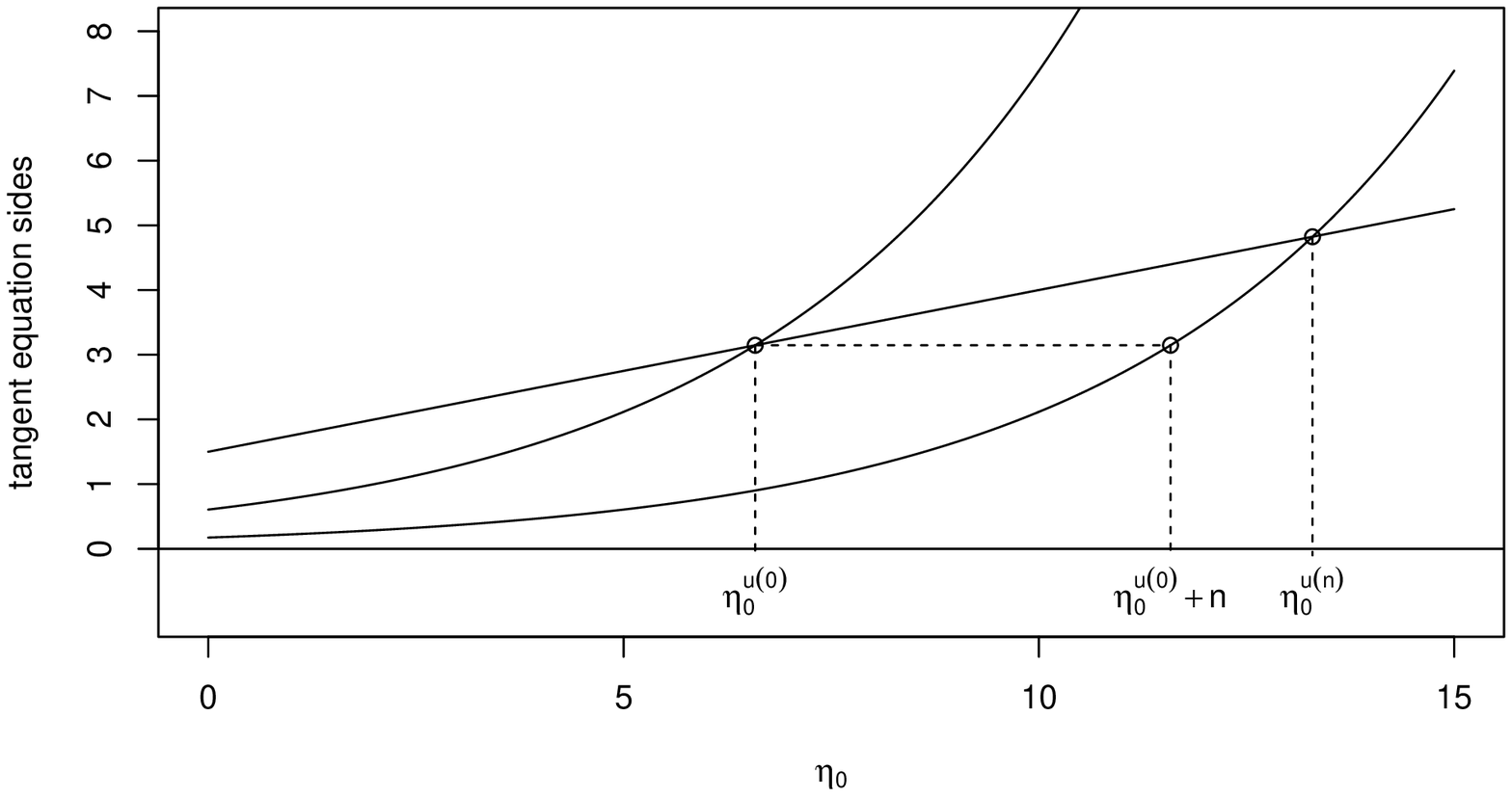}%
%}
\caption{Illustration for the argument that $\ezun > \ezuz + n$.}
\label{fig:spda1}
\end{figure}

\vspace{-0.3cm}
Sets \eqref{eq:basicset}
lead to reduced imprecision in inferences when data are strongly supporting prior information
as the touchpoint moves further to the right in that case. %when updating with strong-agreement data.
The basic shape is symmetric around the $\ez$ axis ($\EZ$ has central ray $y_c = 0.5$),
and updating with strong-agreement data $s/n = 0.5$ means that $\EZ$ is shifted along the $\ez$ axis by $n$,
such that also $\EN$ is symmetric around the $\ez$ axis.
We thus need to consider only one touchpoint.
%With the upper prior touchpoint at $\ezuz$,
Movement to the right means that the upper posterior touchpoint $\ezun$
is larger than the updated prior touchpoint $\ezuz$,
so we need to show that $\ezun > \ezuz + n$.
The upper contour for the posterior boatshape,
updated with $s = \frac{n}{2}$, %has its `prow' now at $(\ezl + n, 0)$,
is $\czu$ from \eqref{eq:basiccontours} shifted to the right by $n$, i.e.,
%\begin{align*}
$\cnu(\eta_0) = a - a e^{-b(\eta_0 - n -\ezl)}$.
%\frac{d}{d\eta_0} \ol{c}(\eta_0) &= ab e^{-b(\eta_0 - n - \ezl)} \,.
%\end{align*}
The equation to identify the posterior upper touchpoint is
\begin{align}
%a - a \big(1 + b({\eta_0^u}\un + 2)\big) e^{-b({\eta_0^u}\un - n - \ezl)} &\stackrel{!}{=} 0 \nonumber\\
1 + b(\ezun + 2) &\stackrel{!}{=} e^{b(\ezun - n - \ezl)} \,.
\label{eq:eta0uposterior}
\end{align}
Comparing \eqref{eq:eta0uposterior} to \eqref{eq:eta0uprior},
%we can conclude that indeed ${\eta_0^u}\un > {\eta_0^u}\uz + n$.
%
both have a linear function with slope $b$ and intercept $1+2b$
on the left hand side.
%the left hand side of both %\eqref{eq:eta0uprior} and \eqref{eq:eta0uposterior}
%gives the identical linear function with slope $b$ and intercept $1+2b$. %of ${\eta_0^u}\uz$ or ${\eta_0^u}\un$, respectively.
The exponential function on the right hand side of \eqref{eq:eta0uposterior} is the function 
on the right hand side of \eqref{eq:eta0uprior} shifted to the right by $n$.
We can picture this situation as in Fig.~\ref{fig:spda1}:
$\ezuz$ is identified by the intersection of the linear function with the left,
non-shifted exponential, %function,
whereas $\ezun$ is at the intersection of the linear function with the right,
shifted exponential. %function.
Since $b > 0$, we have indeed $\ezun > \ezuz + n$.
%${\eta_0^u}\un$ is necessarily larger than ${\eta_0^u}\uz + n$.
%
%We can also conclude that the larger $b$, the steeper the linear function
%and thus the larger the distance of $\ezun$ to $\ezuz + n$,
%that is, the stronger the prior-data agreement effect of the shape.
%Can see this also from normalworld***: a bulkyer the shape has more ueberhang
%which gets reduced by updating as it is located towards the left side of the set***

\vspace{-0.3cm}

%\subsection{Touchpoints for Update with \texorpdfstring{$s > \frac{n}{2}$}{s > n/2}}
\subsection{Touchpoints for Arbitrary Updates}
\label{sec:generalupdate}
\vspace{-0.3cm}

Let us now consider the update of the basic boatshape \eqref{eq:basicset} %symmetric around the $\eta_0$ axis,
in the general case $s \neq \frac{n}{2}$,
investigating the effect that different values of $s$ for fixed $n$ have
on $\ezln$ and $\ezun$
\footnote{We treat $s$ as a a real-value in $[0,n]$ for convenience of our discussions; this does not
affect the conclusions.}
For $s \neq \frac{n}{2}$, $\EN$ is not symmetric to the $\ez$ axis,
and we have to derive the touchpoints $\ezln$ and $\ezun$ separately.
The upper and lower contours for $\EN$ are
\begin{align*}
\ol{c}\un(\eta_0)                   &= s - \frac{n}{2} + a - a e^{-b(\eta_0 - n - \ezl)} \,, &
%\frac{d}{d\eta_0} \ol{c}(\eta_0) &=                      ab e^{-b(\eta_0 - n - \ezl)} \,,\\
\ul{c}\un(\eta_0)                   &= s - \frac{n}{2} - a + a e^{-b(\eta_0 - n - \ezl)} \,,
%\frac{d}{d\eta_0} \ul{c}(\eta_0) &=                     -ab e^{-b(\eta_0 - n - \ezl)} \,.
\end{align*}
leading to
\begin{align}
%s - \frac{n}{2} + a - a \big(1 + b(\eta_0^u + 2)\big) e^{-b(\eta_0^u - n - \ezl)} &\stackrel{!}{=} 0 \nonumber\\
\label{eq:eta0u-general}
\frac{a}{s - \frac{n}{2} + a} \big(1 + b(\ezun + 2)\big) &\stackrel{!}{=} e^{b(\ezun - n - \ezl)} \,,\\
\label{eq:eta0l-general}
\frac{a}{\frac{n}{2} -s  + a} \big(1 + b(\ezln + 2)\big) &\stackrel{!}{=} e^{b(\ezln - n - \ezl)} \,.
\end{align}
We see that the graph from Fig.~\ref{fig:spda1} holds here as well,
except that the linear function on the left hand side of 
\eqref{eq:eta0u-general} and \eqref{eq:eta0l-general} is changed in slope and intercept by a factor.
(Equivalently, we can consider it to be rotated around the root $-2-\frac{1}{b}$.)
For $s=\frac{n}{2}$, this factor is 1 for both \eqref{eq:eta0u-general} and \eqref{eq:eta0l-general},
reducing to \eqref{eq:eta0uposterior}.
Due to symmetry of $\EZ$ we consider, without loss of generality, only the case $s > \frac{n}{2}$.

%\paragraph{Description of ${\eta_0^u}\un$.}

The factor $\frac{a}{s - \frac{n}{2} + a}$
in \eqref{eq:eta0u-general} is smaller than $1$ and decreasing in $s$ to $\frac{a}{\frac{n}{2} + a}$ for $s=n$.
As the linear function's slope will be less steep (the intercept is lowered as well),
%\footnote{The common root of the linear functions for any $s$ is at $\eta_0 = -2 -\frac{1}{b}$.}
the intersection with the exponential function moves to the left,
i.e.\ $\ezun(s) < \ezun(\frac{n}{2})$ for $\frac{n}{2} < s < n$.
This means that $\ynu_\Eta(s) > \ynu_\Eta(\frac{n}{2})$.
However, $\ezun(s)$ can decrease only to $\ezl+n$:
When $\ezun(s)$ reaches the left end of $\EN$ at $\ezl+n$,
the gradual increase of $\ynu_\Eta$ through the changing tangent slope 
%for $\ezl+n \le \eta_0^u(s) \le \eta_0^u(\frac{n}{2})$ 
is replaced by a different change mechanism,
where increase of $\ynu_\Eta$ is solely due to the shift of $\EN$ in the $\eta_1$ coordinate.
Due to \eqref{eq:trafotony}, $\ynu_\Eta$ is then linear in $s$.

%\paragraph{Description of ${\eta_0^l}\un$.}

In \eqref{eq:eta0l-general}, the factor to the linear function is $\frac{a}{\frac{n}{2} - s + a}$.
Here, we have to distinguish the two cases $\frac{n}{2} \le s < \frac{n}{2} + a$
and $s \ge \frac{n}{2} + a$.
In the first case, the factor is larger than $1$ and increasing in $s$ so
the intersection of the linear function with the exponential function
will move to the right, such that $\ezln(s)$ becomes larger, and $\ynl_\Eta$ increases.
In the second case, the factor is undefined (for $s = \frac{n}{2} + a$)
or negative (for $s > \frac{n}{2} + a$) and there is no intersection of the linear function 
with the exponential function for any $\eta_0 > \ezl + n$. So for $s \ge \frac{n}{2} + a$, 
the whole set is above the $\eta_0$ axis,
and the touchpoint must thus be at $\ezu + n$.
Actually, $\ezln(s) = \ezu + n$ already for some $\frac{n}{2} \le s < \frac{n}{2} + a$,
when the intersection point reaches $\ezu + n$.
At this point, gradual increase of $\ynl_\Eta$ resulting from the movement of $\ezln(s)$ along the set
towards the right is replaced by a linear increase in $s$.
Again, this is because the $\eta_1$ coordinate is incremented according to \eqref{eq:eta-update},
and from \eqref{eq:trafotony} we see that $\ynl$ is linear in $\eta_1$.

\vspace{-0.3cm}
\subsection{Posterior Imprecision}
\label{sec:posteriorimprecision}
\vspace{-0.3cm}
We now summarize the results from Sect.~\ref{sec:generalupdate}
and give two numerical examples.
For $s > \frac{n}{2}$, both $\ynu$ and $\ynl$ will at first increase gradually with $s$,
as $\ezun$ moves to the left, and $\ezln$ moves to the right.
We will call such updating of the prior parameter set,
where both lower and upper posterior touchpoints are in the middle of the set, \emph{happy learning}.
\begin{figure}
\includegraphics[width=0.9\textwidth]{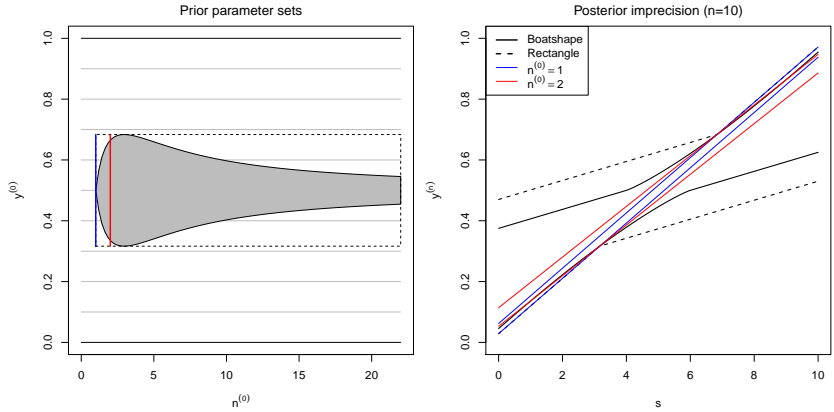}
\caption{Boatshape set with $y_c = 0.5$ together with rectangle set, $1 \times [\yzl, \yzu]$ and $2 \times [\yzl, \yzu]$
with same prior imprecision (left),
and the corresponding lower and upper bounds for $\yn$ as functions of $s$ (right).}
\label{fig:ppp1}
\end{figure}
At some $s^u$, $\ezun$ will reach $\ezl + n$,
and at some $s^l$, $\ezln$ will reach $\ezu + n$.
Whether $s^l < s^u$ or vice versa depends on
the choice of parameters $\ezl, \ezu, a$ and $b$.
When $s$ is larger than either $s^l$ or $s^u$,
we have \emph{unhappy learning},
where data $s$ is very much out of line with our prior expectations as expressed by $\EZ$.
Ultimately, when $s > s^u$ and $s > s^l$,
both $\ynu_\Eta$ and $\ynl_\Eta$ will increase linearly in $s$, but with different slopes.
$\ynu_\Eta$ will increase with slope $\frac{1}{\ezl + n + 2}$,
whereas $\ynl_\Eta$ will increase with the lower slope $\frac{1}{\ezu + n + 2}$.

These findings are illustrated in Fig.~\ref{fig:ppp1}
for a boatshape set with $y_c = 0.5$, $\ezl = -1$, $\ezu = 20$, $a=1$ and $b=0.4$. 
These are compared to a rectangular set and two line segment sets with the same $\yz$ range.
Here we see a linear increase of $\ynu_\Eta$ for $s < 4$ and a superlinear increase for $s \ge 4$.
%and analogously for $\ynl$
We have happy learning for $s \in [4,6]$, and unhappy learning for $s \neq [4,6]$.
For $s \approx 5$, $\Delta_y$ for the boatshape set is about half of $\Delta_y$ for the rectangle set.
The line segment sets lead to very short $\yn$ ranges, but do not reflect \pdc.

\begin{figure}
\includegraphics[width=0.9\textwidth]{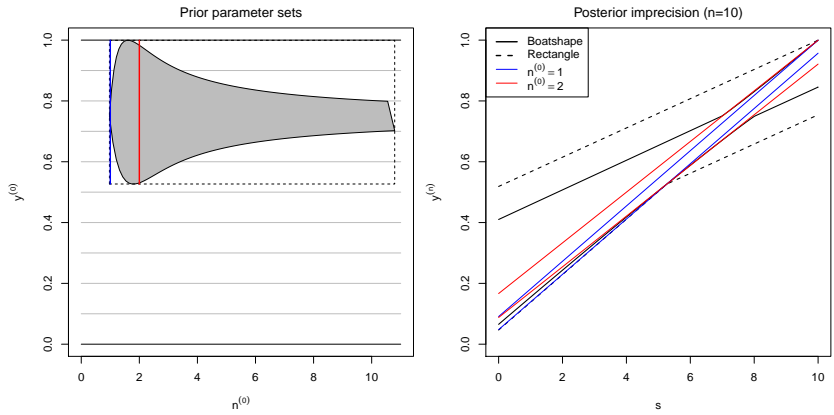}
\caption{Boatshape set with $y_c = 0.75$ together with rectangle set, $1 \times [\yzl, \yzu]$ and $2 \times [\yzl, \yzu]$
with same prior imprecision (left),
and the corresponding lower and upper bounds for $\yn$ as functions of $s$ (right).}
\label{fig:ppp2}
\end{figure}

Figure~\ref{fig:ppp2} depicts a numerical example for the case $y_c = 0.75$.
Notice that the rotated boatshape parameter set is not symmetric in the $(\nz, \yz)$ space.
We see that $[\ynl_\Eta, \ynu_\Eta]$ is nearly as short as $[\ynl, \ynu]$ for the line segments sets
when $s \approx 0.75$, but that unlike those, the boatshape offers \pdc\ sensitivity.
Interestingly, all four sets lead to a similar $\ynl$ for $s < 5$.

\section{Concluding Remarks}
\label{sec:concluding}

%We have described a new method for generating prior parameter sets in a conjugate setting
%that, in addition to prior-data conflict, reflect strong prior-data agreement.
%The parameter set was defined in terms of a novel parametrization \cite{2015:mik-isipta}.
%
For application of the novel method presented in this paper, elicitation of the boatshape 
set parameters must be considered, pre-posterior analysis seems useful for this.
It will be interesting to investigate whether another way of defining a set aligned to a certain ray could be useful,
namely by shifting each part of $\EZ$ from \eqref{eq:basicset} in the $\eo$ dimension
onto the desired ray (similar to turning a right prism into an oblique prism). %according to \eqref{eq:raysofconstantexpectation}.
Alternatives to the functional form of the contour functions \eqref{eq:basiccontours} could also be worth of study.
The method was presented here for the case of binary data,
it can be easily generalized to cover all sample distributions that belong to the exponential family,
since for those a conjugate prior in the $(\ez, \eo)$ parametrization can be constructed
having a purely data-dependent translation as update step \cite[p.~56]{2015:mik-isipta}.

In the parameter space described in Sect.~\ref{sec:miksworld},
updating the prior set
amounts to a purely data-dependent translation, leaving the set shape unchanged.
As shown, this enables flexible modeling of prior information and tailored posterior inference properties,
while remaining within the generalized Bayesian paradigm, hence opening a wide field of research on prior set shapes for specific inference objectives.

\subsubsection{Acknowledgements}
Gero Walter was supported by the Dinalog project
``Coordinated Advanced Maintenance and Logistics Planning for the Process Industries'' (CAMPI).

% ------------ bibliography -------------

%\section*{References}
%\printbibliography[heading=none]
\bibliographystyle{splncs03}
\bibliography{boatpaper}

\end{document}